\documentclass[11pt]{article}
\usepackage{graphicx}
\usepackage{amssymb,amsmath,cite}

\title{Prevention versus treatment: A game-theoretic approach}
\author{Romulus Breban\footnote{Email: romulus.breban@pasteur.fr}\\Unit\'e d'Epid\'emiologie des Maladies Emergentes,\\ Institut Pasteur, Paris, 75724}
\date{\today}
\begin{document}
\maketitle
\begin{abstract}
Empirical studies show that preference for prevention versus treatment remains a subject of debate. We build a paradigm model combining a utility game for the individual-level dilemma of prevention versus treatment, and a compartmental model for the epidemic dynamic. We assume that individuals arrive to maximize the utility of voluntary prevention, as the epidemic reaches an endemic level alleviated by prevention and treatment. We thus obtain an expression for the asymptotic prevention coverage. Notably, we obtain that, if the relative cost of prevention versus treatment is sufficiently low, epidemics may be averted through the use of prevention alone. 
\end{abstract}

\section{Introduction}
Prevention and treatment are complementary strategies for the public health authority. Allocation of limited resources between these strategies remains under continuous debate \cite{Gervas:2008gy,Corso:2006wl}. Benefit-cost analyses for the long and short term may yield conflicting results.  In addition, individual-level perceptions of prevention versus treatment could be biased to jeopardize implementation success of public health programs.  Incentives may be used by the practitioners to redirect individual's attention and alleviate this problem \cite{DeJaegher:2010ci}.

The issue of prevention versus treatment has been addressed in the context of particular medical issues such as HIV/AIDS \cite{Brock:2009cz}, cardiovascular mortality \cite{Ford:2011en}, well-being in work environment \cite{Reynolds:1997ji} and use of ondansetron against intrathecal morphine-induced pruritus \cite{Kung:2014cu}. Surveys studies assessed overall preference for prevention versus treatment in the general population, using discrete choice questionnaires for qualitative and monetary appreciation, as well as propensity to pay \cite{Corso:2002vu,Meertens:2013cy,Luyten:2015da}. The results of these studies show that preference for prevention or treatment depends on the circumstances. A different approach was to present a policy dilemma of prevention versus treatment to a random sample of individuals. Their uninformed opinions compiled for summary statistics, do not show but a slight preference for prevention versus treatment \cite{Johannesson:1997kd,Ubel:1998tf}. Furthermore, a study shows preference for environmental programs, rather than public health programs altogether, in a comparison of four program pairs \cite{Subramanian:2000ky}.  

In the late nineties, Geoffard and Philipson \cite{Geoffard:1997df} opened the theoretical discussion on the setup where prevention is achieved through voluntary vaccination, in absence of treatment. They used an $SIR$ model to describe epidemic dynamics and  assumed that an individual's probability to get vaccinated increases with the perceived utility of vaccination, depending on the probability that a nonvaccinator is infected during his entire lifetime. Geoffard and Philipson focused on pricing strategies for vaccines that provide complete immunity, given the law of supply and demand tuned by disease prevalence. Their main result is that demand on the vaccine private market decreases sufficiently fast toward disease elimination, to make eradication impossible. Bauch and Earn \cite{2004PNAS..10113391B} proposed a different modeling approach, assuming that the individual-level utility is based only on the current state of the epidemic, and that a typical individual arrives to maximize the utility of vaccination as a new endemic level is achieved in the presence of the vaccination program. They also discussed the role of the relative cost of vaccination versus infection morbidity in determining disease prevalence at the endemic state. They found that it is impossible to eradicate a disease through voluntary vaccination when individuals act according to their own interests.  Subsequent literature built upon this modeling framework, including new modeling aspects; see e.g., Refs.~\cite{Bauch:2012yu,Manfredi:2009nr}.

Here, we analyze the theoretical setup where prevention becomes available for a treatable disease with complete natural recovery. We assume that, in the absence of an epidemic, the use of prevention is not motivated at the population scale. We propose a mathematical model to describe the interplay of prevention and treatment, where prevention is voluntary and treatment is mandatory. The model consists of two parts. The epidemic dynamics is described by an $SIR$ model expressed in terms of ordinary differential equations. In addition, we use a utility game to describe the individual-level decisions to join the prevention program. The primary focus of the paper is to discuss how the prevention coverage, reached during an epidemic, depends on the parameters of the prevention and treatment programs. We find that it is possible for prevention programs to avert an epidemic if the relative cost of prevention versus treatment is sufficiently low.

\section{Model}
Our model consists of two ingredients. First, we use a system of ordinary differential equations to mimic epidemic dynamics, not explicitly reported to the public. Second, we specifically describe the individual-level process of making the decision of whether to use  prevention. The primary goal of individuals is to avoid infection and the impact of infection on their everyday activity. Individuals may be aware of means of prevention and treatment, even though they do not not fully understand the corresponding  population-level mechanisms. According to game theory, their decision making skill may be summarized by a utility function  comprising behavioral parameters to describe how individuals perceive the impact of infection on themselves alone.  We assume that the utility function carries no information on the epidemic dynamics, depending solely on the current state of the epidemic.  A fundamental parameter of the utility function could be the current prevalence of infection, as a surrogate for the probability of acquiring infection. This information may be obtained explicitly, through public health reports, or implicitly, through trial, error and adjusting personal decisions to collective opinion. 

Epidemic severity depends on the coverage of prevention and treatment programs and, in turn, motivates individual-level, voluntary decisions to enroll in these programs. We analyze the situation where prevention is made available during the course of an epidemic of a treatable disease and, as a result, the epidemic reaches an endemic level, depending on the prevention coverage. We assume that the endemic state is reached as individuals maximize their perceived utility function for prevention versus treatment. 

\subsection{Compartmental model}
We model an infectious disease with natural recovery.  However, we assume that treatment is available in unlimited supply, so recovered status may also be achieved through treatment.  Although treatment may be offered on a voluntary basis, we assume that all infectious individuals make the choice of being treated according to the regulations of the public health authority. This transmission dynamic may be described by an $SIR$-type model. In addition to treatment, we assume that a prevention program is always in place, whether or not there is an epidemic. We propose the following differential equations
\begin{eqnarray}
\label{eq:P}
\frac{dP}{dt}&=&p\pi-\mu P,\\
\label{eq:S}
\frac{dS}{dt}&=&(1-p)\pi-\mu S-\frac{\beta SI}{N},\\
\label{eq:I}
\frac{dI}{dt}&=&\frac{\beta SI}{N}-(\mu+\nu+\gamma)I,\\
\label{eq:T}
\frac{dT}{dt}&=&\xi\gamma I-\mu T,\\
\label{eq:R}
\frac{dR}{dt}&=&(1-\xi)\gamma I+\nu I-\mu R,
\end{eqnarray}
where the variables stand for: participants to prevention programs $(P)$, susceptible $(S)$, infectious $(I)$ and recovered $(R)$ individuals, as well as individuals cured due to treatment $(T)$; $N=P+S+I+T+R$.  The parameters are as follows. The symbol $p$ stands for the probability of using prevention, $\pi$ for the susceptible inflow, $\mu$ for disease-unrelated death rate, $\beta$ for disease transmissibility, $\nu$ for the natural recovery rate and $\gamma$ for the treatment rate. The symbols $p$ and $\xi$ represent the coverage of the prevention program and treatment efficacy, respectively. 
\indent The model has two equilibria, a disease-free state and an endemic state. The basic reproduction number of the model given by Eqs.~\eqref{eq:P}-\eqref{eq:R} is $(1-p)R_0$, where $R_0\equiv\beta/(\mu+\nu+\gamma)$ is the basic reproduction number of the $SIR$ model without prevention. The key result from the $SIR$ model is the endemic prevalence of infectious individuals, which may be written as 
\begin{eqnarray}
\Pi=\frac{(1-p)-1/R_0}{1+(\nu+\gamma)/\mu},
\end{eqnarray}
where $\Pi>0$ if and only if $R_0(1-p)>1$.

\subsection{Utility game}
Individuals must volunteer to enroll into prevention programs. We model the process of choice as a utility game, including behavioral assumptions for how each individual analyzes pros and cons of prevention versus treatment to make his choice. Individuals make their decisions based on an ensemble of mixed strategies of prevention and treatment. In all, we summarize the utility of prevention versus treatment for a typical individual as follows 
\begin{eqnarray}
U(p,c_p,c_t)=\left\{\begin{array}{ll} 0,&\rm{if~} R_0\leq 1;\\ p\,(-c_p)-(1-p)c_t\Pi, & \rm{if~} R_0>1. \end{array}\right.
\end{eqnarray}
Hence, in the case where $R_0>1$, individuals would pay the cost of prevention $c_p$ with probability that they use prevention $p$. Otherwise, if infected, they pay for the cost of treatment $c_t$.  We assume that individuals act in their own interest to maximize the utility of prevention, as the epidemic evolves toward the endemic level. For further convenience, we use $r$ for $c_p/c_t$, the relative cost of prevention versus treatment. It is reasonable to assume that prevention is more cost-effective than treatment, hence $0\leq r\leq1$. We also use $U(p,r)$ for the rescaled utility function $U(p,c_p,c_t)/c_t$.

\section{Results} 
Maximization of the likelihood $U(p,r)$ yields the following formula for the probability to enroll in prevention programs
\begin{eqnarray}
p(r)=\left\{\begin{array}{ll} 1-\frac{r}{\hat r}-\frac{1}{2R_0}, & {\rm if~} 0\leq r<\hat r -\frac{\hat r}{2R_0};\\ 0, & {\rm if~} \hat r -\frac{\hat r}{2R_0}\leq r;\end{array}\right.
\end{eqnarray}
where 
\begin{eqnarray}
\hat r\equiv\frac{2}{1+(\nu+\gamma)/\mu}.
\end{eqnarray}

\begin{figure}[h!]\begin{center}
\includegraphics[width=0.6\textwidth]{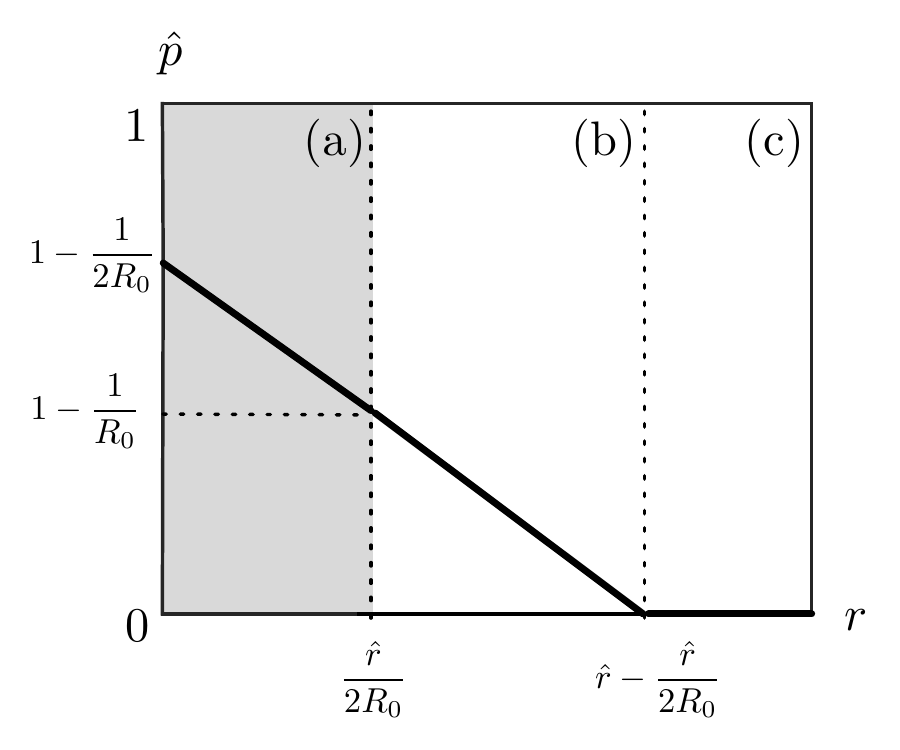}
\caption{Probability of using prevention $p(r)$ as a function of the relative cost of prevention versus treatment $r$. The $[0,1]$ interval on the horizontal axis is divided into three, corresponding to the epidemiological impact of prevention. For region (a), the epidemic is prevented. For region (b), the epidemic is alleviated by prevention. For region (c), individuals do not use prevention measures. We note that the minimum coverage to prevent an epidemic is $p(\hat r/(2R_0))=1-1/R_0$, a well-known result for vaccination models.}\label{fig:1}\end{center}
\end{figure}

From epidemiological perspective, the domain of the function $p(r)$ is divided into three intervals, represented in Fig.~\ref{fig:1} as regions (a), (b) and (c). For region (a), the relative cost of prevention versus treatment $r$ is small, so many enough individuals adopt prevention and the epidemic is averted. Mathematically, we have 
\begin{eqnarray}
0\leq r\leq\hat r/(2R_0)\Leftrightarrow 1/2\leq (1-p)R_0\leq 1.
\end{eqnarray}
We note that the minimum coverage to prevent an epidemic is $p(\hat r/(2R_0))=1-1/R_0$, a well-known result for vaccination models; see Refs.~\cite{Anderson:1991vz}, p.~87 and \cite{Diekmann:2000wp}, Chapter 6.  For region (b), the cost of prevention versus treatment is moderate; i.e., $\hat r/(2R_0)< r<\hat r-\hat r/(2R_0)$. Hence, individuals adopt prevention and the epidemic is alleviated. For region (c), the relative cost of prevention versus treatment is too high and individuals disregard the possibility of disease prevention.
That is, if $\hat r-\hat r/(2R_0)\leq r$, then $p=0$.

This theoretical setup applies for the case of vaccination against childhood diseases.  Parameter values for $\nu$ and $\gamma$ may be found in Ref.~\cite{Anderson:1991vz}, p.~31.  Hence, we have $\hat r/(2R_0)\sim10^{-3}$.  That is, if prevention is thousands of times less costly that treatment, epidemics are averted by prevention alone. We postulate that this is the case for vaccination against childhood diseases.

\section{Discussion and conclusion}
In this paper, we address the impact of voluntary prevention versus mandatory treatment for controlling an epidemic. Previously, Geoffard and Philipson \cite{Geoffard:1997df} argued that, in the case where access to prevention (i.e., vaccination) is determined by disease prevalence and the law of supply and demand in private markets, pricing may render impossible to eliminate the disease through prevention. However, the public health authority may overrule the law of supply and demand by, for example, eliminating copayments for health supplies and services. 

Bauch and Earn \cite{2004PNAS..10113391B} discussed the individual-level dilemma of to prevent or not prevent using vaccination, in the absence of treatment, using a mixed model consisting of ordinary differential equations and a utility game. The individual-level dilemma rested on the notion of herd immunity alone. This is as follows. If an individual uses prevention, he avoids infection. However, when the individual does not use prevention, he may still avoid infection if his peers use prevention in sufficient numbers such that an epidemic does not occur. Bauch and Earn concluded that voluntary prevention cannot drive the epidemic extinct.

Here, we discussed how the prevention coverage depends on the relative cost of prevention versus treatment, when cost does not change because of epidemic severity. The individual-level dilemma that occurs in this case rests on the choice of health strategy. An individual may use prevention to avoid infection starting immediately. Otherwise, he will eventually get infected and acquire permanent immunity upon treatment or natural recovery. Notably, according to our modeling framework, voluntary prevention may prevent epidemics.

Voluntary prevention versus mandatory treatment is a problem ubiquitous in nature. For example, this paradigm question is paramount to economics, for pricing insurance policy. Another application is to military strategy, known as the dilemma of preemptive versus defensive war. In this work, the discussion addressed public health policy.  

\begin{thebibliography}{10}
\bibitem{Gervas:2008gy}
Juan G{\'e}rvas, Barbara Starfield, and Iona Heath.
\newblock {Is clinical prevention better than cure?}
\newblock {\em The Lancet}, 372(9654):1997--1999, December 2008.
\bibitem{Corso:2006wl}
P~S Corso.
\newblock {Prevention just in case or treatment just because: measuring
  societal preferences}.
\newblock {\em Harvard Health Policy Rev}, 2006.
\bibitem{DeJaegher:2010ci}
Kris De~Jaegher.
\newblock {Physician incentives: Cure versus prevention}.
\newblock {\em Journal of Health Economics}, 29(1):124--136, January 2010.
\bibitem{Brock:2009cz}
D~W Brock and D~Wikler.
\newblock {Ethical Challenges In Long-Term Funding For HIV/AIDS}.
\newblock {\em Health Affairs}, 28(6):1666--1676, November 2009.
\bibitem{Ford:2011en}
Earl~S Ford and Simon Capewell.
\newblock {Proportion of the Decline in Cardiovascular Mortality Disease due to
  Prevention Versus Treatment: Public Health Versus Clinical Care}.
\newblock {\em Annual Review of Public Health}, 32(1):5--22, April 2011.
\bibitem{Reynolds:1997ji}
S~Reynolds.
\newblock {Psychological well-being at work: is prevention better than cure?}
\newblock {\em Journal of psychosomatic research}, 43(1):93--102, 1997.
\bibitem{Kung:2014cu}
A~T Kung, X~Yang, Y~Li, A~Vasudevan, S~Pratt, and P~Hess.
\newblock {Prevention versus treatment of intrathecal morphine-induced pruritus
  with ondansetron}.
\newblock {\em International Journal of Obstetric Anesthesia}, 23(3):222--226,
  August 2014.
\bibitem{Corso:2002vu}
Phaedra~S Corso, James~K Hammitt, John~D Graham, Richard~C Dicker, and Sue~J
  Goldie.
\newblock {Assessing Preferences for Prevention versus Treatment Using
  Willingness to Pay}.
\newblock {\em Medical Decision Making}, 22(suppl 1):s92--s101, September 2002.
\bibitem{Meertens:2013cy}
Ree~M Meertens, Vivian~MJ Van~de Gaar, Maitta Spronken, and Nanne~K de~Vries.
\newblock {Prevention praised, cure preferred: results of between-subjects
  experimental studies comparing (monetary) appreciation for preventive and
  curative interventions}.
\newblock {\em BMC Medical Informatics and Decision Making}, 13(1):136,
  December 2013.
\bibitem{Luyten:2015da}
Jeroen Luyten, Roselinde Kessels, Peter Goos, and Philippe Beutels.
\newblock {Public Preferences for Prioritizing Preventive and Curative Health
  Care Interventions: A Discrete Choice Experiment}.
\newblock {\em Value in Health}, 18(2):224--233, March 2015.
\bibitem{Johannesson:1997kd}
M~Johannesson and P~O Johansson.
\newblock {A note on prevention versus cure}.
\newblock {\em Health policy}, 41(3):181--187, 1997.
\bibitem{Ubel:1998tf}
P~Ubel, M~D Spranka, M~L Dekay, and J~Hershey.
\newblock {Public preferences for prevention versus cure}.
\newblock {\em Med Decis Mak}, 18:141--148, 1998.
\bibitem{Subramanian:2000ky}
Uma Subramanian and Maureen Cropper.
\newblock {Public Choices Between Life Saving Programs: The Tradeoff Between
  Qualitative Factors and Lives Saved}.
\newblock {\em Journal of Risk and Uncertainty}, 21(1):117--149, 2000.
\bibitem{Geoffard:1997df}
P~Y Geoffard and T~Philipson.
\newblock {Disease eradication: private versus public vaccination}.
\newblock {\em The American Economic Review}, 87(1):222--230, 1997.
\bibitem{2004PNAS..10113391B}
Chris~T Bauch and David J.~D. Earn.
\newblock {Vaccination and the theory of games.}
\newblock {\em Proceedings of the National Academy of Sciences of the United
  States of America}, 101(36):13391--13394, 2004.
\bibitem{Bauch:2012yu}
Chris~T Bauch and Samit Bhattacharyya.
\newblock Evolutionary game theory and social learning can determine how
  vaccine scares unfold.
\newblock {\em PLoS Comput Biol}, 8(4):e1002452, 2012.
\bibitem{Manfredi:2009nr}
Piero Manfredi, Pompeo~Della Posta, Alberto d'Onofrio, Ernesto Salinelli,
  Francesca Centrone, Claudia Meo, and Piero Poletti.
\newblock Optimal vaccination choice, vaccination games, and rational
  exemption: an appraisal.
\newblock {\em Vaccine}, 28(1):98--109, Dec 2009.
\bibitem{Anderson:1991vz}
Roy~M Anderson and Robert May.
\newblock {\em {Infectious diseases of humans: dynamics and control}}.
\newblock Oxford University Press, New York, 1991.
\bibitem{Diekmann:2000wp}
O~Diekmann and J~A~P Heesterbeek.
\newblock {\em {Mathematical Epidemiology of Infectious Diseases}}.
\newblock Model Building, Analysis and Interpretation. John Wiley {\&} Sons,
  April 2000.
\end{thebibliography}

\end{document}